\documentclass[twocolumn]{aastex63}

\usepackage{threeparttable, tablefootnote} 

\received{}
\revised{}
\accepted{}
\submitjournal{ApJ}

\shorttitle{Coherent events observed by PSP}
\shortauthors{Perrone et al.}

\graphicspath{{./}{figures/}}

\begin{document}

\title{Coherent events at ion scales in the inner Heliosphere: \textit{Parker Solar Probe} observations during the first Encounter}

\correspondingauthor{Denise Perrone}
\email{denise.perrone@asi.it}

\author[0000-0003-1059-4853]{Denise Perrone}
\affiliation{ASI -- Italian Space Agency, via del Politecnico snc, 00133 Rome, Italy}

\author[0000-0002-2152-0115]{Roberto Bruno}
\affiliation{National Institute for Astrophysics, Institute for Space Astrophysics and Planetology, \\
Via del Fosso del Cavaliere 100, 00133 Roma, Italy}

\author[0000-0003-2647-117X]{Raffaella D'Amicis}
\affiliation{National Institute for Astrophysics, Institute for Space Astrophysics and Planetology, \\
Via del Fosso del Cavaliere 100, 00133 Roma, Italy}

\author[0000-0002-6710-8142]{Daniele Telloni}
\affiliation{National Institute for Astrophysics, Astrophysical Observatory of Torino, \\
Via Osservatorio 20, 10025 Pino Torinese, Italy}

\author[0000-0002-7426-7379]{Rossana De Marco}
\affiliation{National Institute for Astrophysics, Institute for Space Astrophysics and Planetology, \\
Via del Fosso del Cavaliere 100, 00133 Roma, Italy}

\author[0000-0002-5365-7546]{Marco Stangalini}
\affiliation{ASI -- Italian Space Agency, via del Politecnico snc, 00133 Rome, Italy}

\author[0000-0002-8399-3268]{Silvia Perri}
\affiliation{Dipartimento di Fisica, Universit\`a della Calabria, Rende, Italy}

\author[0000-0002-7638-1706]{Oreste Pezzi}
\affiliation{Gran Sasso Science Institute, Viale F. Crispi 7, I-67100 L’Aquila, Italy}
\affiliation{INFN/Laboratori Nazionali del Gran Sasso, Via G. Acitelli 22, I-67100 Assergi (AQ), Italy}
\affiliation{Istituto per la Scienza e Tecnologia dei Plasmi, CNR, Via Amendola 122/D, I-70126 Bari, Italy}

\author[0000-0003-3811-2991]{Olga Alexandrova}
\affiliation{LESIA, Observatoire de Paris, Universit\'e PSL, CNRS, Sorbonne Universit\'e, \\
Univ. Paris Diderot, Sorbonne Paris Cit\'e, 5 place Jules Janssen, 92195 Meudon, France}

\author[0000-0002-1989-3596]{Stuart D. Bale}
\affiliation{Space Sciences Laboratory, University of California, Berkeley, CA, USA} 
\affiliation{Physics Department, University of California, Berkeley, CA, USA}
\affiliation{The Blackett Laboratory, Imperial College London, London, UK} 
\affiliation{School of Physics and Astronomy, Queen Mary University of London, London, UK}

\begin{abstract}
\textit{Parker Solar Probe} has shown the ubiquitous presence of strong magnetic field deflections, namely switchbacks, during its first perihelion where it was embedded in a highly Alfv\'enic slow stream. Here, we study the turbulent magnetic fluctuations around ion scales in three intervals characterized by a different switchback activity, identified by the behaviour of the magnetic field radial component, $B_r$. \textit{Quiet} ($B_r$ does not show significant fluctuations), \textit{weak} ($B_r$ has strong fluctuations but no reversals) and \textit{strong} ($B_r$ has full reversals) periods show a different behaviour also for ion quantities and Alfv\'enicity. However, the spectral analysis shows that each stream is characterized by the typical Kolmogorov/Kraichnan power law in the inertial range, followed by a break around the characteristic ion scales. This frequency range is characterized by strong intermittent activity, with the presence of non-compressive coherent structures, such as current sheets and vortex-like structures, and wave packets, identified as ion cyclotron modes. Although, all these intermittent events have been detected in the three periods, they have a different influence in each of them. Current sheets are dominant in the \textit{strong} period, wave packets are the most common in the \textit{quiet} interval; while, in the \textit{weak} period, a mixture of vortices and wave packets is observed.  This work provides an insight into the heating problem in collisionless plasmas, fitting in the context of the new solar missions, and, especially for \textit{Solar Orbiter}, which will allow an accurate magnetic connectivity analysis, to link the presence of different intermittent events to the source region.
\end{abstract}

\keywords{plasmas --- 
turbulence --- solar wind}

\section{Introduction} 
\label{sec:intro}

A puzzling aspect of the fast solar-wind dynamics consists in the empirical evidence that it is hotter than expected for an adiabatic expanding gas~\citep{Marsch82,Lopez86,Freeman88,Hellinger13,Perrone19a,Perrone19b}. Understanding the physical mechanisms of dissipation, and the related heating, in such a turbulent collision-free system, represents nowadays one of the key issues of plasma physics. Moreover, explaining how irreversible heating is accomplished represents a key challenge for thermodynamics in general, since any mechanism in which collisions are not present is lacking the part of the heating process related to the irreversible degradation of information~\citep{Pezzi19,Matthaeus20}.

Spacecraft measurements generally reveal that the solar-wind plasma is in a state of fully-developed turbulence~\citep{Coleman68,Bruno13}. The energy injected by the Sun into the Heliosphere, in the form of large-wavelength fluctuations, e.g. Alfvén waves, is channeled towards short scales through a turbulent cascade until it can be eventually transferred to the plasma particles as heat~\citep{Kiyani09,Alexandrova09,Sahraoui10}. The magnetic power spectrum manifests, at scales corresponding to the inertial range, a behaviour reminiscent of the power-law that characterizes fluid turbulence~\citep{Kolmogorov41,Tu95}. Then, the turbulent cascade extends to smaller scales down to a wavelength range where ions become unmagnetized and the plasma dynamics is governed by particle kinetic properties. At these scales, around ion characteristic lengths, different physical processes come into play, leading to both changes in the spectral shape~\citep{Leamon98,Bale05,Alexandrova13,Lion16} and departure of the ion distribution functions from the thermodynamic equilibrium~\citep{Marsch06,Servidio15,Sorriso19}. 

Turbulence in the solar wind is strongly space-localized and the degree of non-homogeneity increases as the spatial scales decrease. This features, called intermittency and recovered in both magnetic field and plasma parameters~\citep{Veltri99,Bruno03,Salem09,Bruno19}, has been observed to evolve with distance from the Sun and it is due to the presence of coherent structures~\citep{Bruno03,Greco12a}, which can be described as strong non-homogeneities in the magnetic field~\citep{Retino07,Perri12,Greco14,Perrone16,Perrone17} over a wide range of scales~\citep{Greco16,Lion16}. Near these coherent structures, particle energization, temperature anisotropy, and strong deviation from Maxwellian have been observed in both \textit{in situ} data and numerical simulations~\citep{Matteini10,Osman11,Greco12b,Servidio12,Servidio15,Servidio17,Wu13,Perrone13,Perrone14,Pezzi18,Sorriso18}.

Recently, a statistical study of coherent structures at ion scales has been performed in both slow~\citep{Perrone16} and fast~\citep{Perrone17} solar wind at $1$~au by using \textit{Cluster} observations. It has been shown, for the first time, that different families of structures characterize the ion scales of the turbulent cascade of solar-wind plasma. This means that different mechanisms of dissipation occur at ion scales, since different structures interact with particles in different ways. Compressive structures, such as magnetic holes, solitons and shocks, and Alfv\'enic structures, in form of current sheets and vortices, are observed in slow solar wind~\citep{Perrone16}. In fast solar wind it has been found that the ion scales are dominated by Alfv\'en vortices, with a small and/or finite compressive part. Current sheets are also observed but no compressive structures are found~\citep{Perrone17}. In this respect, slow wind presents a more complex physics with respect to fast wind, where in the former a significant percentage of structures moves in the flow, maybe leading to the generation of instabilities with additional effects on particles~\citep{Hellinger19}. 

\begin{figure*} 
          \centering
 	\includegraphics[width=17.8cm]{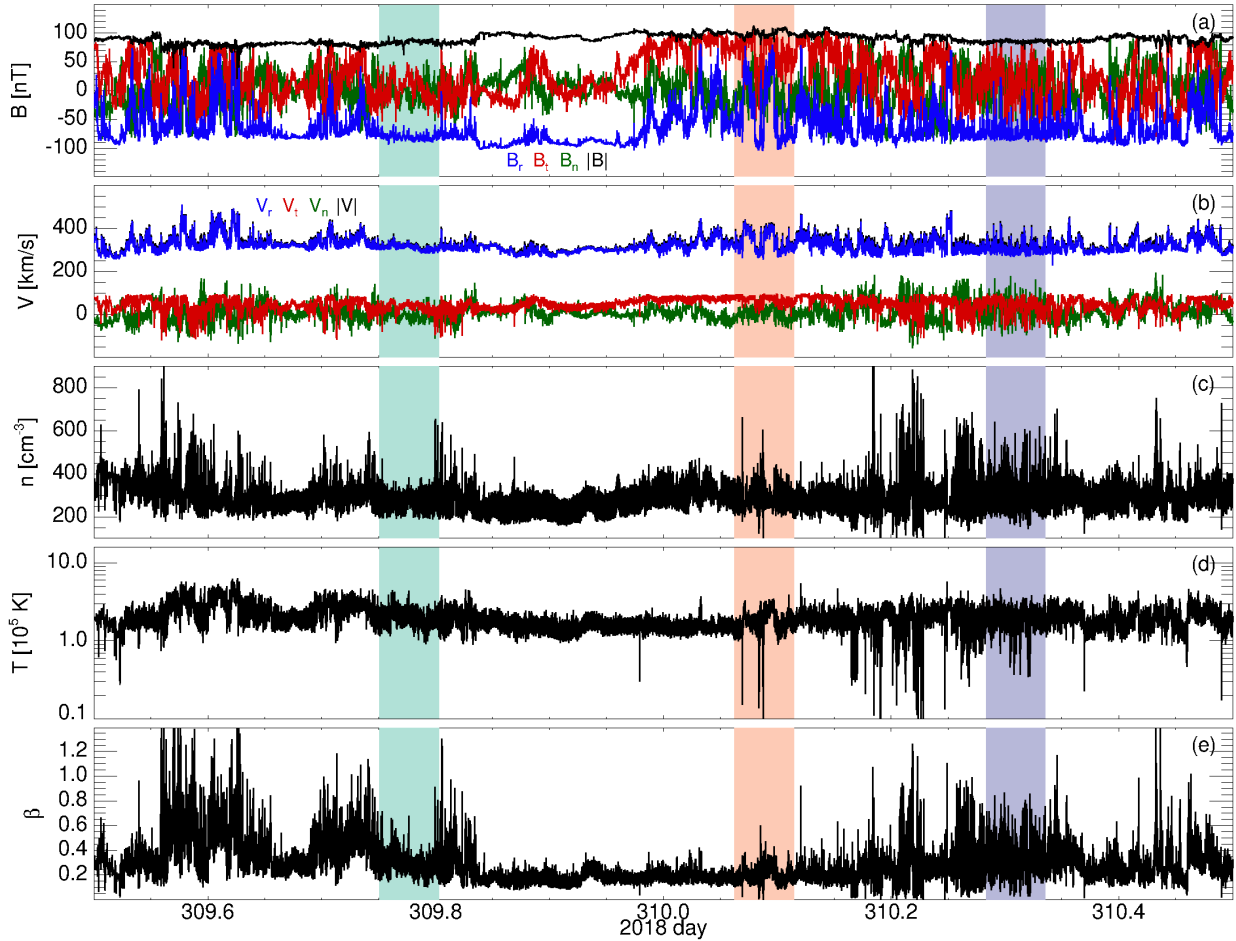}
         \caption{Overview of solar wind data during the first perihelion of \textit{Parker Solar Probe} at $\sim 0.17$~au. From top to bottom: three components in RTN (radial in blue, tangential in red and normal in green) and magnitude (in black) of the magnetic (a) and velocity (b) field vectors; ion density (c) temperature (d), and plasma beta (e). The coloured bands denote three different periods, lasting 1h15, of \textit{quiet} (green), \textit{weak} (violet), and \textit{strong} (orange) switchback activity.}
         \label{fig:stream}
 \end{figure*}

A large variety of magnetic structures has also been detected in the inner heliosphere by using \textit{Messenger} magnetic field observations at $0.3$~au~\citep{Greco14} during the minimum of solar cycle 23. Unfortunately, due to the unavailability of the plasma data on the spacecraft, no information about the stream in terms of origin and/or speed is possible. However, by looking at magnetic field fluctuations, both Alfv\'enic, such as rotational and tangential discontinuties, and compressive structures, namely shocks and magnetic holes, have been identified. The study of the presence of coherent structures in the inner heliosphere, especially in regions close to the Sun, could help to explore the dynamical development of solar wind turbulence. Thanks to \textit{Parker Solar Probe}~\citep[PSP,][]{Fox16}, launched in August 2018, it is possible to study the presence and, eventually, the nature of coherent structures in a completely unexplored environment.

During its first perihelion, \textit{PSP} revealed the presence of a large number of S-shaped magnetic structures which produce patches of large, intermittent magnetic field reversals, namely `switchbacks'~\citep{Bale19,Dudok20}, heat flux reversals and isolated intermittent velocity enhancements, namely `spikes'~\citep{Kasper19,Horbury20}. It is worth to remark that magnetic switchbacks have already been observed at different radial distances from the Sun by previous missions~\citep{Behannon81,Tsurutani94,Kahler96,Balogh99,Landi06,Borovsky16} but always in fast wind. Using \textit{PSP} data, \citet{Krasnoselskikh20} performed a detailed analysis on three selected structures, representative of three different groups of switchbacks, namely (i) Alfv\'enic-like structures, (ii) compressional-like structures, and (iii) full reversals of the radial component of the magnetic field vector. The size of these structures is large compared to the typical characteristic ion scales, except for their boundaries where flowing currents were found. Moreover, a rather intense wave activity, close to the edges of these structures has also been observed. This analysis suggests that these structures are localized twisted magnetic tubes moving with respect to the surrounding plasma.  

 \begin{table*}
 	\centering
 	\caption{Typical solar wind parameters, averaged within each 1h15 selected period, observed by \textit{PSP} at 0.17~au. Day and UT information refers to the starting time.}
 	\label{tab:intervals}
 	\begin{tabular}{ccccccccccccc}
 		\hline
        Period & Day & UT & 
 				        $B$ [nT] & $V$ [km/s] & $n$ [cm$^{-3}]$ & $T$ [10$^5$K] & $\beta$ & $V_A$ [km/s] &
 				        $\lambda_i$ [km] & $\rho_i$ [km] & $\Omega_{ci}$ [rad/s] \\
 		\hline
 		\hline
		Quiet  & 309 & 18:14:24 & 84.3 & 312.2 & 
 			301.1 & 1.9 & 0.28 & 106.0 & 12.1 & 6.9 & 8.1 \\
 		Weak   & 310 & 06:47:31 & 85.6 & 311.4 & 
 			331.2 & 2.0 & 0.32 & 102.6 & 12.5 & 7.0 & 8.2 \\
 		Strong & 310 & 01:29:17 & 99.5 & 344.8 & 
 			306.1 & 1.8 & 0.19 & 124.0 & 13.0 & 5.6 & 9.5 \\
 		\hline		
 	\end{tabular}
 \end{table*}

In this paper, we study the nature of the turbulent magnetic fluctuations around ion scales in three \textit{PSP} intervals with different characteristics during its first perihelion. In particular, by looking at the radial component of the magnetic field vector, $B_r$, we selected 1h15 intervals where (i) $B_r$ is almost constant, (ii) $B_r$ has strong fluctuations but no reversals, and (iii) $B_r$ has full reversals. In these three intervals, we focus on intermittency and we statistically study the observed coherent events. Examples of these events are also shown below.  

The paper is organized as follows: in Section~\ref{sec:PSP} we present the main characteristics of the first \textit{PSP} encounter and we select the three intervals for the analysis; in Section~\ref{sec:struct} we show the results of statistical studies and we present examples of different coherent events; and in Section~\ref{sec:discuss} we summarize the results and discuss our conclusions.

\section{First PSP Encounter} 
\label{sec:PSP}

In November 2018 \textit{PSP} completed its first perihelion passage at about 0.17~au from the Sun, sampling the solar wind far closer than before. During this first encounter, the spacecraft was almost corotating with the Sun and observed a long interval of slow Alfv\'enic wind originating from a small equatorial coronal hole~\citep{Bale19,Badman20}. These observations have shown the presence of isolated intermittent velocity enhancements~\citep{Kasper19,Horbury20} associated with magnetic field deflections~\citep{Bale19,Dudok20}.

In this paper, we consider a one-day period (from 12:00 UT on 2018 November 5) around perihelion (see Figure~\ref{fig:stream}) to study the nature of the turbulent magnetic fluctuations in three 1h15 intervals, with different characteristics in terms of switchbacks. We mainly focus on magnetic field measurements from the fluxgate Magnetometer (MAG), which is part of the FIELDS suite~\citep{Bale16}. We use the full cadence observations of the three components of the magnetic field vector, having a sampling rate of 293~Hz. 
Moreover, in order to set a context of the environment, we consider the solar wind bulk plasma properties measured by the Solar Probe Cup~\citep[SPC,][]{Case20}, a solar-facing Faraday cup, from the SWEAP instrument suite~\citep{Kasper16}. The ion density, $n$, and velocity, ${\bf V}$, are derived by taking moments of the one-dimensional ion velocity distribution obtained by the current spectra, the primary data product of the SPC sensor, with a cadence of about 0.87~s. The ion temperature, $T$, is also estimated from the ion thermal speed, $v_{th}$, which is also a regular derived product of SPC, as $T=m v_{th}^2/2k_B$, being $m$ the proton mass and $k_B$ the Boltzmann's constant.

\subsection{Intervals Characterization}
\label{subsec:intervals}

In Figure~\ref{fig:stream}, an overview of the solar wind in the considered one-day period around the first perihelion of \textit{PSP} is summarized, where magnetic field data have been downsampled to comply with the resolution of particle measurements. Panels (a) and (b) show the three components in RTN reference frame (radial in blue, tangential in red and normal in green) and magnitude (in black) of the magnetic and velocity field vectors, respectively. It is clear, especially looking at the radial component of the magnetic field vector, that different regimes lie in this one-day interval, where the switchback activity significantly varies. In particular, we observe periods where $B_r$ is almost constant and periods where $B_r$ has strong fluctuations and sometimes fully reverses. Therefore, we decided to separately perform our studies on three 1h15 selected periods, denoted in Figure~\ref{fig:stream} by coloured bands. From now on, we will refer to \textit{quiet} (green band), \textit{weak} (violet band) and \textit{strong} (orange band) periods if $B_r$ does not show significant fluctuations, $B_r$ has strong fluctuations but no reversals and $B_r$ has full reversals, respectively. Panels (c) and (d) display the ion density and temperature, respectively. Differences are also recovered in these quantities with respect to the three selected periods. In particular, the largest fluctuations around a mean value are found in the \textit{weak} interval, while in the \textit{strong} one a change in the behaviour is observed in correspondence of the reversals. Finally, in panel (e), we show the ion plasma beta, $\beta$, defined as the ratio between ion kinetic and magnetic pressures. Also in this case, we found the same features observed for the density. 

\begin{figure*}
          \centering
 	\includegraphics[width=17.8cm]{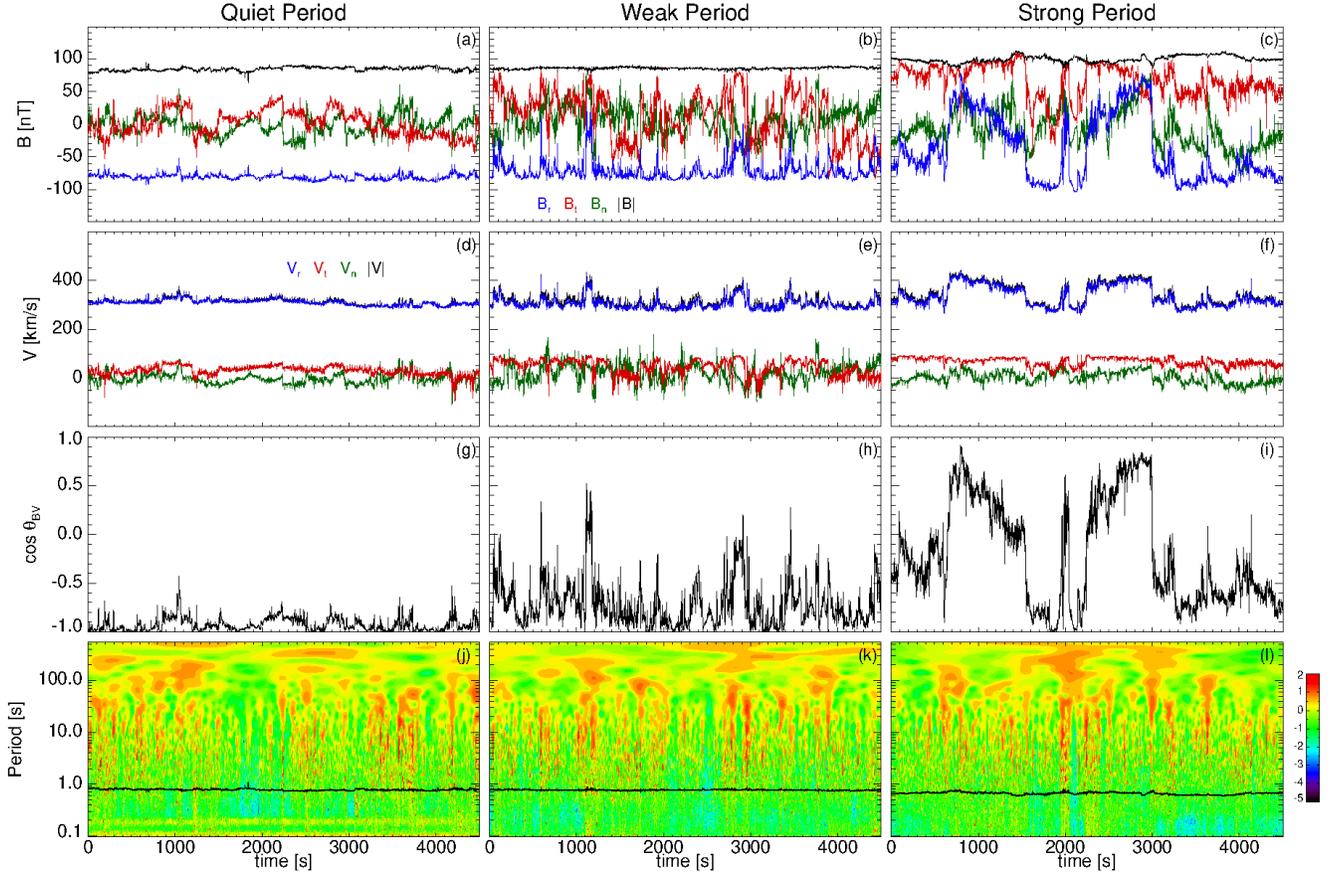}
         \caption{Characteristics of the \textit{quiet} (left column), \textit{weak} (middle column), and \textit{strong} (right column) periods. From top to bottom: zoom of Figure~\ref{fig:stream} in the selected intervals for the magnetic (first row) and velocity (second row) field vectors; cosine of the angle between the magnetic and velocity field vectors (third row); logarithmic contour plots of the Local Intermittency Measure (LIM) of the total magnetic field fluctuations (fourth row). In the latter, black lines denote the ion cyclotron timescale.}
         \label{fig:intervals}
 \end{figure*}
 
Looking at the typical solar-wind parameters (listed in Table~\ref{tab:intervals}), averaged within each selected interval considered in this study, we observe that \textit{quiet} and \textit{weak} periods have almost the same values of the magnitude of the magnetic and velocity field vectors, while in the \textit{strong} period higher values are found. The same trend is observed for the characteristic derived quantities, such as the Alfv\'en speed, $V_A$, the ion Larmor radius, $\rho_i$, and the ion cyclotron frequency, $\Omega_{ci}$. Moreover, we see that the ion density is larger for the \textit{weak} period, probably due to the presence of stronger fluctuations, while the temperature and the ion skin depth, $\lambda_i$, have similar values in all the three intervals. Finally, the ion plasma beta is lower in the \textit{strong} period ($\beta \sim 0.19$), while is higher in the \textit{weak} interval ($\beta \sim 0.32$).

In Figure~\ref{fig:intervals}, we look more in detail at the three periods. The first two rows are just a zoom of Figure~\ref{fig:stream} in the selected intervals for the magnetic and velocity field vectors, respectively. Here, we can better appreciate the behaviour of $B_r$: in the \textit{quiet} (a) period it is almost constant, in the \textit{weak} (b) period it undergoes short scale variations, and in the \textit{strong} (c) period it largely rotates. Moreover, \textit{quiet} (a) and \textit{weak} (b) periods have almost constant magnetic field magnitude, while $B$ in the \textit{strong} period (c) is more variable. Furthermore, for the velocity field vector, we observed an increase of spikes in $V_r$ for the \textit{weak} (e) and \textit{strong} (f) intervals with respect to the \textit{quiet} one (d). Finally, the $V_r$ component is highly correlated with $B_r$ in all the three periods, which of course indicates a high level of Alfv\'enicity in the wind. To better stress this aspect, we also plot the cosine of the angle between ${\bf B}$ and ${\bf V}$ (third row). The \textit{quiet} period shows a very high degree of Alfv\'enicity, with anti-parallel magnetic and velocity field vectors (g). Lower Alfv\'enicity is found in both \textit{weak} (h) and \textit{strong} (i) intervals, even if we can recognize that the change in sign follows the fluctuations of the magnetic field. 

The last row of Figure~\ref{fig:intervals} shows the  Local Intermittency Measure~\citep[LIM,][]{Farge92} for the total magnetic field fluctuations, $|\mathcal{W_{\bf B}}(\tau,t)|^2 = \sum_i |\mathcal{W}_i(\tau,t)|^2$ with $i=r, t, n$, defined as 
\begin{eqnarray}
I(\tau,t) = \frac{|\mathcal{W_{\bf B}}(\tau,t)|^2}{\langle |\mathcal{W_{\bf B}}(\tau,t)|^2 \rangle_t}
\end{eqnarray}
where the brackets indicate a time average and $\mathcal{W}_i(\tau,t)$ are the Morlet wavelet coefficients for different timescales $\tau$ and time $t$~\citep{Torrence98}
\begin{eqnarray}
\mathcal{W}_i(\tau,t) = \sum_{j=0}^{N-1} B_i(t_j)\psi^*\left[ \left( t_j-t \right)/\tau \right] \ .
\end{eqnarray}
Black lines denote the local ion cyclotron timescale. In all the three intervals, the distribution of energy in time and scales (not affected by edge effects) is not uniform, with the appearance of localized energetic events covering a certain range of scales, which are easily recognized by the red color. This is an indication of the presence of coherent structures in the system which emerge at larger time scales and are connected through the scales~\citep{Perrone16,Perrone17,Lion16,Alexandrova20}. Similar features have also been reported by~\citet{Greco16} using the partial variance of increments technique. In panels (j)--(l), we can also observe that sometimes the energy appears localized around the ion characteristic scales (see, e.g., the energy appearing after $t=1000$~s in panel (k) around the inverse of the ion cyclotron frequency). This would suggest the localized presence of some wave activity~\citep{Alexandrova04,Lion16,Bowen20}.

 \begin{figure}
          \centering
 	\includegraphics[width=8.5cm]{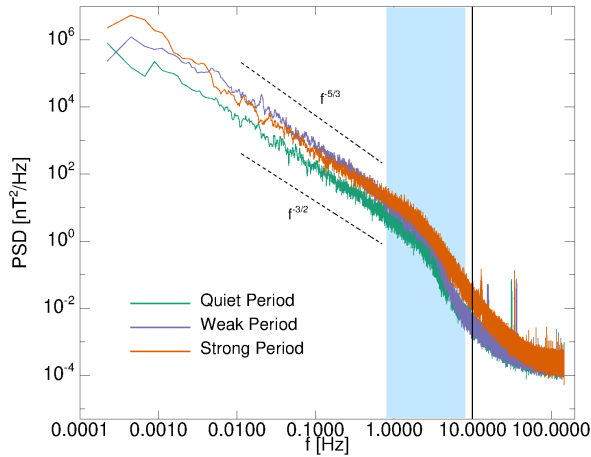}
         \caption{Power Spectral Density (PSD) of the total magnetic field fluctuations for the \textit{quiet} (green), \textit{weak} (violet), and \textit{strong} (orange) periods. Both Kolmogorov and Kraichnan scalings (black dotted lines) have been plotted for reference. The vertical black solid line indicates the maximum resolved frequency for the spectra ($f_{max} = 10$~Hz), while the blue filled band denotes the range of scales, $f\in[0.8,8]$~Hz, around ion scales.} 
         \label{fig:spectra}
 \end{figure}

Figure~\ref{fig:spectra} shows the total power spectral density (i.e. the trace of the spectral matrix) of the magnetic field for the \textit{quiet} (green), \textit{weak} (violet) and \textit{strong} (orange) periods. The vertical solid line at $f = 10$~Hz indicates the frequency at which the noise level becomes significant. These spectra show the characteristic behaviour of the solar wind turbulent cascade. For each stream, we observe at low frequencies a power-law trend between the Kolmogorov~\citep{Kolmogorov41} and the Kraichnan~\citep{Kraichnan65} scaling (black dotted lines). Indeed, in the inertial range, in the frequency range $f \in [0.016, 0.8]$~Hz, the spectral indices for the \textit{quiet}, \textit{weak} and \textit{strong} intervals are $-1.548 \pm 0.015$, $-1.759 \pm 0.014$, and $-1.487 \pm 0.013$, respectively. These values are in agreements with the results described in \citet{Chen20} and \citet{Duan20}. Then, a break between the inertial range and the dissipative range of the turbulent cascade can be identified around ion scales. The frequency location of the break for each time interval, which is around 2 and 3 Hz, has been estimated adopting the same procedure described in~\citet{Bruno14}, closer to the prediction based on the ion cyclotron resonance mechanism~\citep{Leamon98}, in agreement with previous studies \citep{Leamon98, Bruno14, Woodham18, Damicis19, Duan20}. In particular, for the three different time intervals, \textit{quiet}, \textit{weak} and \textit{strong}, we obtain the following ion cyclotron resonance frequencies: $2.49$~Hz, $2.53$~Hz and $2.95$~Hz, respectively, considering that the ion cyclotron frequency is, for the three intervals, 1.29 Hz, 1.30 Hz and 1.51 Hz. To compare, both Doppler shifted ion Larmor radius and inertia length can be estimated, under the assumption of wave vectors parallel to the plasma flow, by using the information in Table~\ref{tab:intervals}. Indeed, we found $f_{\lambda_i}\simeq 3.8$~Hz, $4$~Hz, $4.2$~Hz and $f_{\rho_i}\simeq 7.2$~Hz, $7.1$~Hz, $9.8$~Hz, for \textit{quiet}, \textit{weak} and \textit{strong} interval, respectively.

For the present study we are interested in the investigation of the nature of the turbulent fluctuations around ion scales, indicated in Figure~\ref{fig:spectra} by the blue filled band. Therefore, from now on, our analysis will focus on the denoted range of scales $f \in [0.8,8]$~Hz.

\begin{figure*}
          \centering
 	\includegraphics[width=13.8cm]{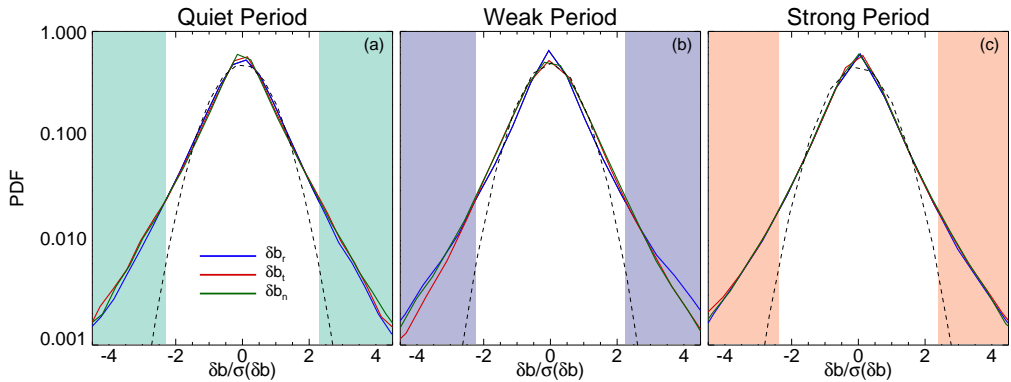}
         \caption{Probability distribution functions (PDFs) of $\delta b_r$ (blue), $\delta b_t$ (red), and $\delta b_n$ (green), normalized to their own standard deviations, $\sigma(\delta b_i)$, for the \textit{quiet} (a), \textit{weak} (b), and \textit{strong} (c) periods. In each panel, black dashed lines, correspond to a Gaussian fit and the filled bands show the regions where magnetic fluctuations exceed three standard deviations of each Gaussian fit.}
         \label{fig:PDF}
 \end{figure*}

\section{Coherent structures}
\label{sec:struct}

In order to look at the overall nature of the magnetic field fluctuations in the chosen range of scales, we adopt a bandpass filter based on the wavelet transform~\citep{Torrence98,He12,Roberts13,Perrone16,Perrone17}. The fluctuations are defined as
\begin{eqnarray}
\label{eq:fluct}
    \delta b_i(t) = \frac{\delta j \delta t^{1/2}}{C_\delta \psi_0(0)} \sum_{j=j_1}^{j_2} \frac{\mathcal{R} \left[ \mathcal{W}_i (\tau_j,t)\right]}{\tau_j^{1/2}}
\end{eqnarray}
where $\mathcal{R}$ refers to the real-part function, $j$ is the scale index and $\delta j$ is the constant scale step; $\psi_0(0)=\pi^{1/4}$ and $C_\delta=0.776$, the latter derived from the reconstruction of a $\delta$ function using the Morlet wavelet~\citep{Torrence98}. Finally, $\tau(j_1)=0.125$~s and $\tau(j_2)=1.25$~s (being $\tau =1/f$).

In Figure~\ref{fig:PDF}, we show the probability distribution functions (PDFs) of the three components of the magnetic field fluctuations ($\delta b_r$ in blue, $\delta b_t$ in red, and $\delta b_n$ in green), normalized to their own standard deviation, $\sigma(\delta b_i)$, for the \textit{quiet} (a), \textit{weak} (b) and \textit{strong} (c) intervals. By comparing the PDFs with a Gaussian fit (black dashed line), we observe the presence of significant non-Gaussian tails in each component of the magnetic field fluctuations, due to the presence of strong energetic events~\citep{Perrone16,Perrone17}. This result is confirmed by the flatness of the magnetic fluctuations, which has the same behavior in the three periods analyzed (not shown). We find that the flatness departs from the normal distribution value around the ion scales, showing an almost flat curve around 4.9, 5.2 and 5.4 for \textit{quiet}, \textit{weak} and \textit{strong} intervals, respectively. This reflects a non-homogeneous distribution of the turbulent fluctuations in all the three periods. In each panels of Figure~\ref{fig:PDF}, the filled bands show the regions where magnetic fluctuations exceed three standard deviations of the corresponding Gaussian fit, which include $99.7\%$ of the Gaussian contribution. We will use the corresponding value in each period as a threshold to select non-Gaussian intermittent events. More than thousand events have been detected in each period, thus supporting a statistical analysis of their properties.

\subsection{Statistical Analysis}
\label{subsec:stat}

\begin{figure}
          \centering
 	\includegraphics[width=7.8cm]{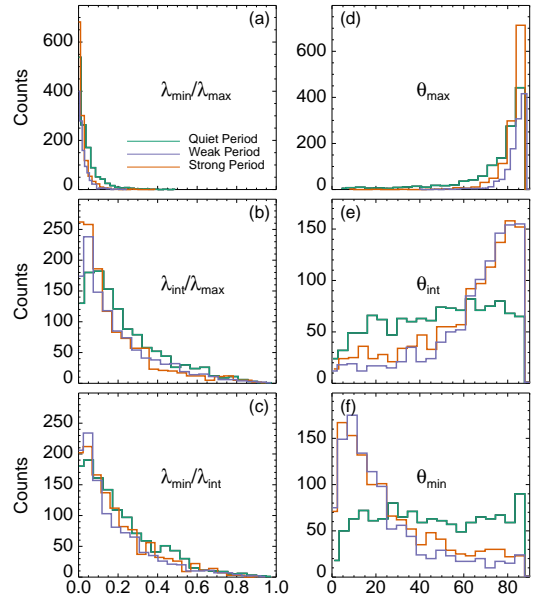}
         \caption{Statistical analysis of the observed coherent events in the minimum variance reference frame for the \textit{quiet} (green), \textit{weak} (violet), and \textit{strong} (orange) periods. Left panels: histograms for the minimum (a) and intermediate (b) eigenvalues normalized to the maximum eigenvalue, and for $\lambda_{min}/\lambda_{int}$ (c). Right panels: histograms of the angles between the maximum (d), intermediate (e), and minimum (f) variance directions and the local magnetic field.}
         \label{fig:MVRF}
 \end{figure}

Figure~\ref{fig:MVRF} shows the results of the minimum variance analysis~\citep{Sonnerup98} applied to the intermittent events observed in the \textit{quiet} (green), \textit{weak} (violet), and \textit{strong} (orange) periods. It is worth pointing out that this analysis has been performed around each selected peak in magnetic energy, which identifies magnetic fluctuations well-localized in time and with regular profiles, in a defined time range, $\Delta t^{'}$, between two minima of energy. This corresponds to the width (i.e. extension) of an event, larger than its characteristic temporal scale, $\Delta t$, which is defined as the width at half height~\citep[see][for more details on the identification method for intermittent events]{Perrone16}. 

The left column displays the histograms for the minimum, $\lambda_{min}$ (a) and intermediate, $\lambda_{int}$ (b) eigenvalues normalized to the maximum eigenvalue, $\lambda_{max}$, and for $\lambda_{min}/\lambda_{int}$ (c). Although most of the considered events, in all the intervals, have $\lambda_{min} \ll \lambda_{int} \ll \lambda_{max}$, we also find the presence of fluctuations with $\lambda_{min} \ll \lambda_{int} \lesssim \lambda_{max}$, a feature a bit more pronounced in the \textit{quiet} period. In general, the minimum variance direction is well defined in all the three considered periods, even if for a very few events, a degeneracy $\lambda_{min} \sim \lambda_{int}$ exists.

The right column shows the histograms of the orientation of the eigenvectors with respect to the local magnetic field, ${\bf b}_0$, also averaged within the structure, thus in $\Delta t^{'}$. The direction of maximum variation, $\theta_{max}$ (d), is perpendicular to ${\bf b}_0$ in all the three intervals, suggesting the absence of compressive events. Differences between the periods, instead, are found in the distributions of $\theta_{int}$ (e) and $\theta_{min}$ (f). In particular, we observe in the \textit{quiet} period an almost uniform distribution for both $\theta_{int}$ and $\theta_{min}$, while in the \textit{weak} and \textit{strong} intervals $\theta_{int}$ is peaked around $90^\circ$ and $\theta_{min}$ is almost parallel to ${\bf b}_0$.

\begin{figure*}
          \centering
 	\includegraphics[width=15.8cm]{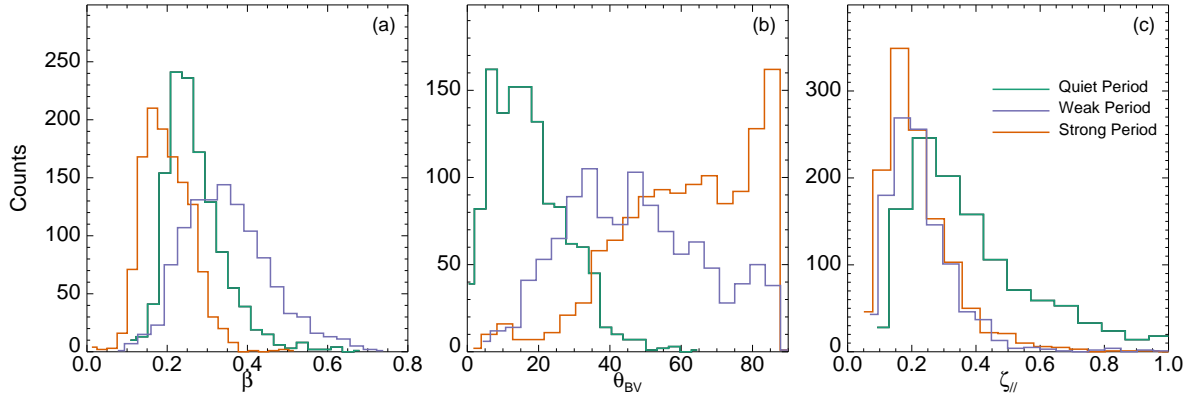}
         \caption{Histograms of $\beta$ (a), $\theta_{BV}$(b), and $\zeta_{\parallel}$ (c), for the observed coherent events in the \textit{quiet} (green), \textit{weak} (violet), and \textit{strong} (orange) periods.}
         \label{fig:stat}
 \end{figure*}

To better highlight the differences between the three intervals with respect to the intermittent events observed, in Figure~\ref{fig:stat} we show the histograms of $\beta$ (a), $\theta_{BV}$ (b), and $\zeta_{\parallel}$ (c), being the latter the local magnetic compressibility~\citep{Perrone16}, defined as 
\begin{eqnarray}
\zeta_\parallel = \sqrt{\frac{\text{max}\left( \delta b_\parallel^2\right)}{\text{max}\left( \delta b_{\perp1}^2+\delta b_{\perp2}^2\right)}}
\end{eqnarray}
where parallel and perpendicular directions are evaluated with respect to ${\bf b}_0$ and the maximum of the magnetic components is evaluated within $\Delta t^{'}$; while $\beta$ and $\theta_{BV}$ have been evaluated as mean values in the same time range. The distribution of the ion plasma beta displays three distinct peaks, which reflects the behaviour of $\beta$ in the whole three periods (see Table~\ref{tab:intervals}), i.e. the largest value is found for the events in the \textit{weak} interval and the lowest for the ones in the \textit{strong} period. Three distinct peaks are also recovered for the angle between the magnetic and velocity field vectors in each event, where $\theta_{BV} \sim 20^\circ$ for the \textit{quiet} interval (green line), $30^\circ < \theta_{BV} < 50^\circ$ for the \textit{weak} period (violet line), and $\theta_{BV} \sim 90^\circ$ for the \textit{strong} one (orange line). This behaviour suggests that the disalignment of ${\bf b}_0$ and ${\bf v}_0$ increases as the switchback activity increases. It is worth pointing out that the value of $\theta_{BV}$ gives a view of the plasma wave vectors that we are able to measure. In particular, for $\theta_{BV} \sim 0^\circ$ $k_\parallel$ are well measured, while for $\theta_{BV} \sim 90^\circ$ $k_\perp$ turbulence is observed. An opposite behaviour with respect to $\theta_{BV}$, is observed, even if less marked, for the local magnetic compressibility (see Figure~\ref{fig:stat}c). Finally, the distribution of the characteristic time scale of these events, $\Delta t$, in all the three periods, is peaked at about $0.25$~s (not shown), which corresponds to $\sim 0.3$--$0.4$ ion cyclotron timescale or, by assuming that the frozen-in, Taylor hypothesis is fulfilled~\citep{Perri17} to a spatial scale of $\sim 6$--$7$ $\lambda_i$ or $\sim 11$--$15$ $\rho_i$.

\subsection{Examples of Observed Coherent Events}
\label{subsec:exemples}

\begin{figure*}
          \centering
 	\includegraphics[width=17.2cm]{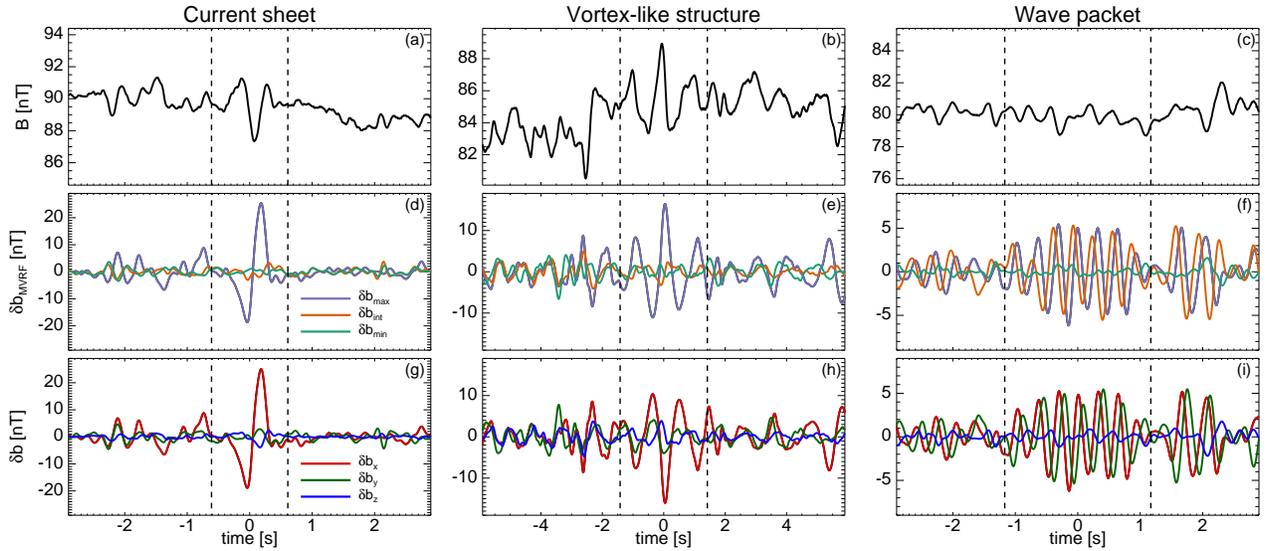}
         \caption{Example of current sheet (left column), vortex-like structure (middle column), and wave packet (right column). Top row: magnitude of the large-scale magnetic field. Middle row: components of the magnetic field fluctuations, defined in Equation~\ref{eq:fluct}, in the minimum variance frame. The maximum direction is in violet, the intermediate in orange and the minimum in green. Bottom row: components of the magnetic field fluctuations in the local magnetic field reference frame, where ${\bf b_0}$ is along the $z$-direction. The vertical black dashed lines mark the extension of the events, $\Delta t^{'}$.}
         \label{fig:examples}
 \end{figure*}

A detailed analysis of all the events observed in the selected periods has allowed us to identify three different families of intermittent events, all of them present in the three periods, namely (i) current sheets, (ii) vortex-like structures, and (iii) wave packets. Examples are shown in Figure~\ref{fig:examples}. The top row displays the modulus of the raw magnetic field, i.e. the large-scale magnetic field as observed by \textit{PSP}, where we have only taken off the noise for $f > 10$~Hz. The three panels have the same aspect ratio to highlight the magnetic compressibility of the events. Middle and bottom rows show the magnetic field fluctuations, as defined in Equation~\ref{eq:fluct}, in the minimum variance reference frame (MVRF) and in the local magnetic field reference frame where ${\bf b_0}$ is along the $z$-direction (${\bf e_z}={\bf e_b}$), $x$ is perpendicular to ${\bf b_0}$ in the plane spanned by it and the radial direction (${\bf e_x}={\bf e_b} \times {\bf e_b}$), and $y$ closes the right-hand reference frame (${\bf e_y}={\bf e_b} \times {\bf e_x})$, respectively. Finally, vertical dashed lines mark the width of the events, $\Delta t^{'}$, where all the analyses have been performed. 

The first example of intermittent event is a current sheet (left column), an incompressible one-dimensional (i.e. linearly polarized) structure, with $\delta b_x$ (g), perpendicular to ${\bf b}_0$, which changes sign. The other two components show very small fluctuations. Moreover, the three components are almost zero in the middle of the structure, where $\delta b_{max}$ reverses (d) and the large-scale magnetic field has a local minimum (a). Finally, the minimum variance analysis shows that the direction of maximum variation is perpendicular to ${\bf b}_0$, with $\theta_{max} \simeq 85^\circ$.

The second example (middle column) suggests the presence of an Alfv\'en vortex~\citep{Alexandrova06,Alexandrova08,Roberts16,Lion16,Perrone17,Wang19}. The large-scale magnetic field shows a modulated fluctuation with a local maximum in the middle of the structure (b). The fluctuations are localized, with the main variation in the direction perpendicular to ${\bf b}_0$, $\delta b_x$ (h). Indeed, from the minimum variance analysis, we found that the maximum variance is perpendicular to the local magnetic field, $\theta_{max} \simeq 87^\circ$, while  $\theta_{int} \simeq 10^\circ$ and $\theta_{min} \simeq 80^\circ$. In the case of a vortex, as we suppose here, where $\lambda_{int} \simeq \lambda_{min}$, the direction of minimum variance corresponds to the normal direction, while the direction of intermediate variance, which is parallel to the local magnetic field, corresponds to the vortex axis. 

The last example (right column) has no a clear behaviour in the large-scale magnetic field (c), but $\delta b_i$ show quasi-monochromatic magnetic fluctuations from $-6$ to $6$~nT, in the plane perpendicular to ${\bf b}_0$ (i). The minimum variance analysis gives $\theta_{max} \simeq 86^\circ$ and $\theta_{int} \simeq 85^\circ$, while the minimum variance direction is almost parallel, $\theta_{min} \simeq 6^\circ$. We also note that the transverse components are out of phase of about $\pi/2$. The observed features can be associated to a wave activity. 

\begin{figure*}
          \centering
 	\includegraphics[width=17.2cm]{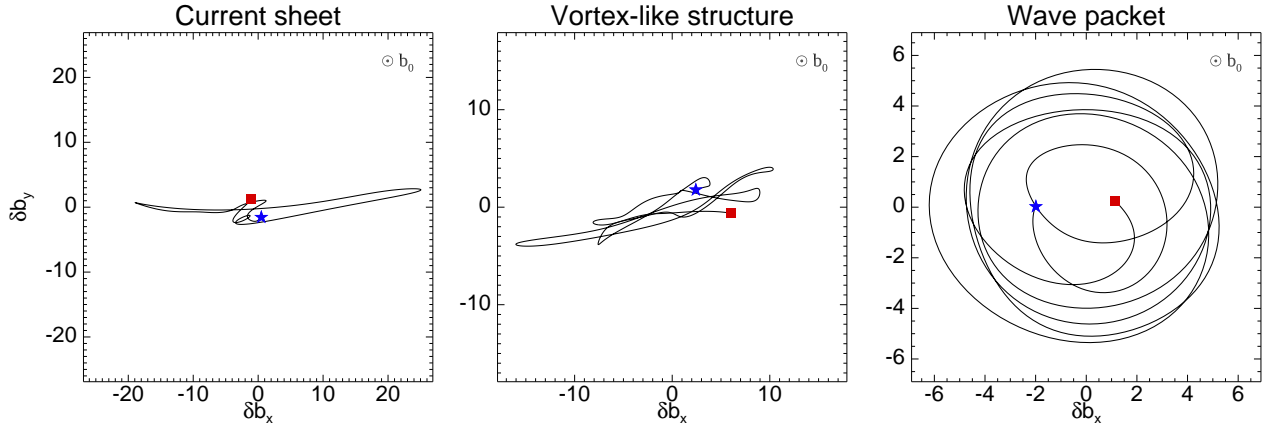}
         \caption{Hodogram for the magnetic field fluctuations of the current sheet (left), vortex-like structure (middle) and wave packet (right) shown in  Figure~\ref{fig:examples}, in the plane perpendicular to the local ${\bf b_0}$. Blue star refers to the starting time point and red square to the end time point.}
         \label{fig:hodogram}
 \end{figure*}

To better highlight the polarization of these events, in Figure~\ref{fig:hodogram} we show the hodograms for the components of the magnetic field fluctuations in the plane perpendicular to the local magnetic field (with the local magnetic field direction out of plane) for the examples of current sheet (left), vortex-like structure (middle) and wave packet (right) shown in Figure~\ref{fig:examples}. Blue star indicates the starting point, while the red square the end one for the magnetic temporal signal. The first two hodograms confirm the linear polarization of both the current sheet and the vortex-like structure. Such linear polarisation for vortices can be found while crossing the vortex very close to its centre~\citep[see e.g.][]{AlexandrovaSaur08}. On the other hand, the fluctuations of the wave packet are clearly left-handed circularly polarized around the direction of ${\bf b_0}$, which is inward directed,  and can be interpreted as outward propagating ion cyclotron waves~\citep{He11,Podesta11,Telloni15}. This is in agreement with the results in~\citet{Huang20}. Ion cyclotron waves are indeed left-handed polarized waves, with a wavevector nearly aligned to the local magnetic field and frequencies around the proton gyrofrequency. In solar wind, they can be found in individual wave packets lasting a few minutes~\citep{Jian09,Lion16,Bowen20} or in `storms' lasting many hours~\citep{Jian10,Jian14,Wicks16,Lion16PHD,Bowen20}. Their presence has also been found in numerical simulations \citep{Pezzi17}. Ion cyclotron waves have recently been directly observed in the solar wind in periods characterized by strong Alfv\'enic fluctuations at inertial scales~\citep[see][and references therein]{Telloni19}.

\section{Discussions} 
\label{sec:discuss}

We have studied in detail the nature of magnetic turbulent fluctuations around the ion characteristic scales in the inner heliosphere, by using the unique opportunity offered by \textit{PSP} which is sampling the solar wind far closer than ever before. In particular, we focused on a one-day interval around its first perihelion, at 0.17~au, where we selected three 1h15 periods characterized by a different switchback activity (looking at the behaviour of $B_r$), and we studied magnetic properties at ion scales, thus smaller than the ones considered in \citet{Krasnoselskikh20}.

The three chosen intervals, namely \textit{quiet} ($B_r$ does not show significant fluctuations), \textit{weak} ($B_r$ has strong fluctuations but no reversals) and \textit{strong} ($B_r$ displays full reversals), show different characteristics also in terms of large-scale plasma quantities. In particular, stronger fluctuations, in both density and temperature, are found for the whole \textit{weak} interval, while in the \textit{strong} one reversals drive changes in their behaviour. Moreover, we observed an increase of the presence of spikes in $V_r$ for the \textit{weak} and \textit{strong} intervals, with respect to the \textit{quiet} period. Furthermore, the magnitude of both the velocity and magnetic field vectors are the same in the \textit{quiet} and \textit{weak} periods, with an almost constant trend, but becomes larger, and with small changes, in the \textit{strong} interval. Finally, differences are observed also for the Alfv\'enicity, which is very strong in the \textit{quiet} interval, while is lower in both \textit{weak} and \textit{strong} periods. It is worth pointing out that, for turbulent fluctuations at the scales considered in this work, the angle between magnetic and velocity field vectors also indicates which wave vectors are observed. Indeed, if $\theta_{BV} \sim 0^\circ$ or $\theta_{BV} \sim 180^\circ$, the satellite is able to scan $k_\parallel$, while for oblique angles $k_\perp$ turbulence is well measured. In this study, $k_\parallel$ is mostly resolved in the \textit{quiet} period, while in the other two intervals $\theta_{BV}$ covers all possible angles, meaning that both $k_\parallel$ and $k_\perp$ can be measured.

The study of the spectral properties for the three considered periods showed that each stream is characterized, in the inertial range, by a power law between the  Kolmogorov spectrum and the Kraichnan scaling. Then a break, around the characteristic ion scales has been observed. In addition, the frequency break location seems to match the one predicted by the ion cyclotron resonance mechanism, confirming previous results based on different s/c observations~\citep{Bruno14, Woodham18, Damicis19, Duan20}. Moreover, we also looked at the distribution of the magnetic energy in time and frequency and we found that localized regions in time that cover a certain range of scales exist in all the considered intervals. We also observed that sometimes energy appears localized around the ion cyclotron frequency. This kind of non-homogeneity in the magnetic energy distribution has already been observed in both slow and fast wind at 1~au and it has been related to the presence, at ion scales, of different families of coherent structures or waves~\citep{Alexandrova04,Lion16,Perrone16,Perrone17}. Motivated by these results, we decided to study magnetic fluctuations in the range $f \in [0.8,8]$~Hz, around the typical ion scales for these intervals, which are also characterized by a significant departure from Gaussianity.

We have detected more than thousand intermittent events in each period, which have well-localized fluctuations in space with regular profiles. These events can be divided in three different families, namely current sheets, vortex-like structures, and wave packets, with different influence in each considered period. A minimum variance analysis has shown that the peak of the distributions for the eigenvalues is found for $\lambda_{min} \ll \lambda_{int} \ll \lambda_{max}$, meaning a prevalence of one-dimensional fluctuations. This is the case, for example, of the current sheets, which are the most common events in the \textit{strong} period ($\sim 46\%$ out of $\sim 220$ events for which the magnetic profile is clear). However, also two-dimensional fluctuations are found, where $\lambda_{int} \lesssim \lambda_{max}$. In particular, values of $\lambda_{int}/\lambda_{max} \in [0.2,0.6]$ can be recovered in case of vortices, while $\lambda_{int} \simeq \lambda_{max}$ can be found for the case of wave packets. In the \textit{weak} period we observe a mixture of vortices ($\sim 45\%$ out of $\sim 342$) and wave packets ($\sim 50\%$), while in the \textit{quiet} period, where $\theta_{BV} \sim 0^\circ$ allowing to resolve $k_\parallel$ waves~\citep{Lion16PHD,Bowen20}, wave packets are the most frequent class of intermittent events ($\sim 61\%$ out of $\sim 303$). The left-handed circular polarization around the direction of the local magnetic field of these wave packets suggests that these wave modes can correspond to ion cyclotron waves. Evidence for the presence of ion cyclotron waves in highly Alfv\'enic periods supports previous findings by~\citet{Telloni19,Telloni20}, where a clear link between the Alfv\'enicity at fluid scales and the existence of ion cyclotron waves at kinetic scales has statistically been proved. Moreover, ion cyclotron waves are associated to high levels of temperature anisotropy, which lead the proton velocity distribution to depart from thermodynamic equilibrium, thus triggering the development of proton cyclotron plasma instability~\citep[see e.g.][]{Gary94,Bourouaine10,Telloni19}. 

The minimum variance analysis has also shown that the direction of maximum variation is, in all the three intervals, perpendicular to the local magnetic field, suggesting the absence of compressive events.  This is in disagreement with the results in slow wind at 1~au~\citep{Perrone16}; however, the slow wind observed by \textit{PSP} is highly Alfv\'enic, with much more similarities with the fast wind~\citep{Damicis19,Stansby20,Perrone20,Telloni20}. In fact, we found a good agreement, for the distribution of $\theta_{max}$ and $\theta_{min}$ in the \textit{quiet} period, with the observations in fast wind at 1~au~\citep{Perrone17}. In particular, in the \textit{quiet} interval we found an almost uniform distribution for both $\theta_{int}$ and $\theta_{min}$, suggesting that the presence of vortices ($\sim 31 \%$), jointly with waves (the dominant contribution), could generate a mix-up of the intermediate and minimum directions, thus all possible angle can be covered. Finally, the very low magnetic compressibility found in our analysis is also in agreement with the results in the inner heliosphere, using \textit{Helios} data in fast solar wind~\citep{Bavassano82}.

Understanding the role of small-scale coherent structures and waves into the general problem of dissipation, and thus heating, in collisionless plasmas, and especially in solar wind, represents a key problem for space plasma physics. \textit{In situ} measurements~\citep{Marsch06,Bourouaine10,Wu13} and numerical experiments~\citep{Araneda08,Valentini08,Perrone13,Perrone14} have shown that particle heating and acceleration and temperature anisotropy appear localized in and near coherent structures or due to wave-particle interactions. Unfortunately, the resolution of the particle measurements on \textit{PSP} is not enough to study in details the kinetics within the selected events (the mean width of the events is around 2~s and the resolutions for ion data is about 1~s). However, some indications can be obtained. In particular, peaks of both ion density and temperature are noticed at the edges of both current sheets and vortex-like structures (not shown). Conversely, we saw a peak within wave packets for both ion density and temperature (not shown).  

The new observations of these macroscopic magnetic switchbacks made by \textit{PSP} are allowing to add new pieces to the puzzle of the energy dissipation mechanisms in collisionless plasmas. However, the origin of such structures remains still unclear and debated. Until now, several physical processes, both in-situ \citep{Squire20, Ruffolo20} and in the solar atmosphere \citep[see for instance][]{Tenerani20}, have been proposed to explain these switchbacks. Among the many possibilities, it was pointed out that these could be due to coronal jets and filament eruptions~\citep{Sterling20}, to reconnection processes or to phenomena happening in the deep corona. It has been shown, through state-of-the-art numerical modelling \citep{Tenerani20}, that these perturbations may indeed originate in the solar atmosphere and propagate upwards. In this regard, it is worth recalling that large amplitude kink-like oscillations are nowadays detected in small-scale magnetic elements at all heights in the solar atmosphere, from the corona and chromospheric heights~\citep[see for instance][to mention a few]{Anfinogentov13, Jafarzadeh17} down to the photosphere, where they are excited by the forcing action of the granular buffeting~\citep{Stangalini14}. In the near future, thanks to the synergy between \textit{PSP}, which will collect measurements far closer to the Sun, and \textit{Solar Orbiter}~\citep{Muller13}, which will combine both remote sensing and \textit{in situ} measurements, it will be possible to provide further insights in the understanding the link between the switchbacks in the solar wind and their possible source in the solar atmosphere.


\end{document}